\documentclass[dvips]{article}
\usepackage{icrctc07}

\title{ANTARES alternative event reconstruction strategies}

\authors{Y. Becherini$^{1}$ on behalf of the ANTARES Collaboration$^{2}$}
\afiliations{$^{1}$ DSM/Dapnia, CEA/Saclay, Yvonne.Becherini@cea.fr \\ $^{2}$ http://antares.in2p3.fr}

\abstract{
The ANTARES Collaboration is building a high-energy neutrino telescope at 2500 m 
depth in the Mediterranean Sea. The experiment aims to search for high-energy 
cosmic neutrinos through the detection of Cerenkov light induced by 
muons and showers resulting from neutrino interactions with the surrounding 
medium. The detector will consist of a three-dimensional array of 900 optical 
modules housing photomultipliers. It will be composed of 12 strings, 5 of them 
being already in operation since January 2007. 
The muon track is reconstructed from the arrival time and the charge of the 
signals obtained from the photomultipliers, whose positions are known by 
means of an acoustic positioning system. The reconstruction strategies include 
several steps among which there are: optical background filtering, algorithms 
for first estimations of the track parameters, and a final fit aiming to reach an 
angular resolution better than 0.3 degree above 10 TeV in the full detector.
Different reconstruction strategies will be presented and their application to the 
present real data analysis will be reviewed.}

\begin{document}
\maketitle
\section{Introduction}

When an ultra-relativistic particle ($\beta \simeq 1$) moves in a medium, Cerenkov light is emitted 
at an angle depending on the refraction index $n$ of the medium. 
In case of sea water $n$ is about $1.34$, and thus the Cerenkov emission angle becomes
\begin{equation}
cos (\theta_C) = \frac{1}{n} \simeq cos(42^\circ).
\end{equation}
The emitted Cerenkov photons are detected by an array of photomultipliers installed at the bottom of the sea.
Since January 2007 the ANTARES detector is a full three-dimensional array of photomultipliers consisting of 5 strings 
detecting muons at a depth of 2475 meters below the sea level. 
ANTARES 10-inch photomultipliers are housed in pressure resistant glass spheres called optical modules (OM).
The 5-line detector data taking has allowed the Collaboration to tune the various processes needed to collect data, 
and muon events have been detected on top of the optical background (60-100 kHz most of the time).
The calibration system has been proven to be efficient, and the knowledge of the charge, of the arrival time of the signals 
and the positioning of the OMs has allowed the first reconstruction codes to be tested in realistic conditions.
The status of the experiment is discussed in \cite{antoine}.
Preliminary data studies show that a flux of atmospheric downward-going muons is triggering the detector at a rate of about 1 Hz
and that upgoing atmospheric neutrino candidates have been identified.
Among the various muon reconstruction algorithms under study, two different alternative methods are presented in this paper.
Other well developed reconstruction methods are described in \cite{aart}, \cite{carmona}, and 
the implementation of the {\sl{simulated annealing}} algorithm \cite{simann} is under study.
A discussion on event reconstruction techniques for Cerenkov neutrino telescopes can also be found in \cite{amanda}.

\section{Track reconstruction strategies} 

\begin{figure}
\begin{center}
\noindent
\includegraphics [width=0.47\textwidth]{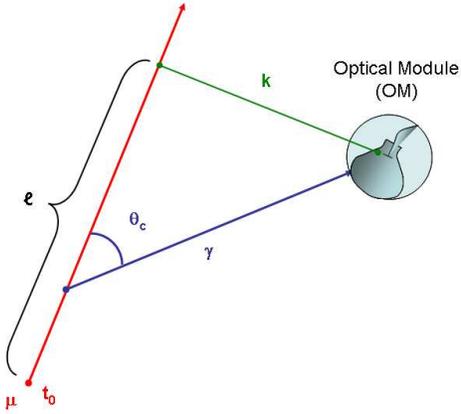}
\end{center}
\caption{\small{Schematic view of the kinematics related to the photon arrival time on the OM. 
The red line represents the muon, while the blue line represents the photon.}}
\label{fig0}
\end{figure}

A minimum-ionising muon crossing the ANTARES detector causes the emission of Cerenkov photons, at an angle $\theta_C$ of about 
$42^\circ$ with respect to the track. 
As the muon energy increases, secondary particles are created by the muon energy loss along its path.
These secondaries produce photons which are delayed with respect to the light produced by the passage of the muon itself.
Photons may also scatter in water, inducing an additional delay of the arrival times on the OMs.  
Assuming that the muon propagates as a straight line at speed $c$,
the arrival time of the direct Cerenkov photon $t^{th}_{i}$, as shown in Fig. \ref{fig0} can be directly linked to the muon track 
through the following equation:
\begin{equation}
t^{th}_{i} = t_0 + \frac{1}{c} \left( \ell - \frac{k}{tan \theta_C} \right) + \frac{1}{v_g} \left( \frac{k}{{\textrm sin}\theta_C} \right)
\label{exp_time}
\end{equation}
where $v_g$ is the group velocity of light, and $\ell$ and $k$ are distances 
depending on the track parameters $\theta, \phi, x, y, z,$ at $t = t_0$, where $t_0$ is the time of the event. 
The track reconstruction methods usually consist in two main steps: a pre-fitting procedure where a first estimation of the track parameters is
performed, and a final fit usually being a maximum likelihood method, where a probability density function (PDF) tries to include the full knowledge 
of the detector and the expected physics. This PDF is a function of the hit time residuals, defined as 
the arrival time differences between the observed hit times and 
the hit times expected for a direct Cerenkov photon travelling directly from the muon to the OM
\begin {equation}
t^{res}_{i} \equiv \frac{t_{i}-t_{i}^{th}}{\sigma}
\label{reseq}
\end{equation}
$t_{i}$ being the measured time, $t_{i}^{th}$ given by Eq. \ref{exp_time} for a given OM position $x_i,y_i,z_i$, 
and $\sigma$ being the time resolution.
A set of estimators, hit selections, fits and pre-fits is called a reconstruction strategy.


\section{Examples of reconstruction strategies}

Two strategies chosen as an illustration of the event reconstruction in ANTARES are presented here. 
The first strategy presented (here called strategy 1) is composed by a set of linear pre-fits, 
and each pre-fit is than taken as first guess for a final $L_1-L_2$ estimator fit as defined in \cite{zhang}.
The second strategy (here called strategy 2 or ScanFit) is composed by an iterative procedure giving a set of 
preferred directions, each selected direction is then fitted with a minimum $\chi^{2}$ method.

\subsection{Pre-fitting procedures}

In general, the pre-fit step in track reconstruction must give an initial guess for the second minimisation step.
The challenge is to find a relatively fast procedure giving a solution which is not too far away from the reality. 
Two pre-fitting procedures will be discussed in the following section: the linear and the scanning pre-fits. 

\subsubsection{Linear pre-fit}

The \textsl{line} or \textsl{Stenger} fit algorithm \cite{stenger} produces an initial estimation of the track parameters 
with the time and amplitude of the hits. 
This algorithm is efficient when the length of the track in the detector is large compared to the light attenuation length.
The locations of each OM ${\bf r}_i$, recording a hit at time $t_i$ can be connected by a line
\begin{equation}
{\bf r}_i \simeq {\bf r} + {\bf v} \cdot t_i
\end{equation}
where {\bf r} is the track position at $t=0$ and {\boldmath ${v}$} the track velocity vector.
Under this assumption, a linear fit using space and time information of the event is performed through
the minimisation of
\begin{equation}
\chi^2 = \sum_{i=1}^{n_{h}} a_i \; ({\bf r}_i - {\bf r} - {\bf v} \cdot t_i)^2
\label{linear_prefit}
\end{equation}
where $n_{h}$ is the number of hits, $a_i$ is the measured charge for the hit $i$ at position {\boldmath $r$}$_i$ 
seen at time $t_i$. The factor $a_i$ takes into account that a hit closer to the track will have a higher amplitude. 
The $\chi^2$ minimum is obtained from the parameter values {\boldmath $r$} and {\boldmath $v$} yielding vanishing partial derivatives.
The resulting Eq. \ref{linear_prefit} can be solved analitically, giving:
\begin{equation}
{\bf r} = \left \langle {\boldmath r}_i \right \rangle - {\bf v} \cdot \left \langle t_i \right \rangle 
\end{equation}
and
\begin{equation}
{\bf v} = \frac{\left \langle  {\bf r}_i \cdot t_i \right \rangle - \left \langle {\bf r}_i \right \rangle \cdot \left \langle t_i \right \rangle} {\left \langle t_i^2 \right \rangle - \left \langle t_i \right \rangle^2}
\end{equation}
where 
\begin{equation}
\left \langle x_i \right \rangle \equiv \frac{1}{n_{h}} \sum_{i=1}^{n_{h}} a_i \; x_i
\end{equation}
is the weighted mean of parameter $x_i$ with respect to all hits.
The linear fit gives a vertex point ${\bf r}$ and a direction ${\bf{u} = \bf{v/|v|}}$ at $t=0$, where $\bf{|v|}$ is the mean speed of light 
propagating through the one-dimensional vector projection.  
Zenith $\theta$ and azimuth $\phi$ angles of the fitted track are finally computed through the direction vector 
$\bf {v}$ and the ($x,y,z,\theta, \phi, t=0$) parameters become the starting point for the subsequent 
minimisation step.
The algorithm is applied to all hits in the event, and once the linear pre-fit solution is found, the track position is shifted to the plane perpendicular to the track 
and containing the barycentre of the detector giving a new position $\bf{r_0}$. The time of the track is calculated accordingly: $t_0 = \bf{r_0}/c$. 
Then the average $\overline{t}$ and the RMS of the hit time residuals with respect to the calculated track are evaluated.
The average of this distribution is added to the track time in order to better match the Cerenkov model: $t_0 \rightarrow t_0 + \overline{t}$.
If the RMS of the time residuals distribution is more than 40 ns, the process is re-iterated and stopped when the condition is satisfied. 
The linear fit gives a angular error of about $10^\circ$ from the true solution for the 12-string detector.  

\subsubsection{Scanning pre-fit}

Another possible way of finding the first guess for the track is an iterative search for the direction $\theta$ and $\phi$ maximising the number 
of hits in the given direction which are close to the assumed track.
Given the set of coincidences in the event, i.e. the hits $i, j$ firing different OMs but the same floor and for which $t_i - t_j$ is less than 20 ns,
the set of hits composed by the earliest hits of the sample can be defined.
For a given set of directions a 1D clustering algorithm \cite{mieke} i.e. a causality relation to determine space time correlations between the hits 
is applied to the set of earliest hits. 
Then, the same algorithm is applied a second time in order to add some other hits belonging to the event.
The goal of this hit selection is to have a sample of hits having a high purity of direct Cerenkov light. 
The resulting set of hits is used to determine the three
remaining coordinates x, y, z, using the full Hesse matrix.
Only candidate directions with at least 5 hits are kept and passed
on to the next stage, together with the selected hits.
This procedure is more accurate than the linear pre-fit and is capable of giving a first guess which deviates only by about $5^\circ$ 
from the true direction for the 12-string detector. 
The disadvantage is that this scanning procedure is computing time consuming.
A similar iterative pre-fit was already discussed in \cite{carmona}.
 
\subsection{Final fitting procedures}

From the first guess obtained by the pre-fit algorithm, a likelihood function $\mathcal{L}$ is used to get a better estimate of the track parameters. 
This is normally done by the minimisation of $-log(\mathcal{L})$.
In general, the likelihood function is the product of the PDFs $\mathcal{L}_i$ computed for each hit i
\begin{equation}
\mathcal{L} = \prod_{i=1}^{n_{h}} \mathcal{L}_i.
\end{equation}
The challenge is to find the best description of $\mathcal{L}_i$ given an initial set of hits.
Normally, the hit time residuals follow a complicated distribution, and a sophisticated parameterisation of the hit arrival time
distribution is needed. Implementing a PDF in case of ANTARES requires a detailed knowledge of the detector response 
and of the photon propagation in water. 
The two strategies described don't make use of a specific PDF parameterisation but choose to use an approximated description of the PDF.
The methods used to minimise are SIMPLEX and MIGRAD for the first strategy and MIGRAD only for the second strategy 
(both can be found in the MINUIT package \cite{minuit}).

\subsubsection{Fit with a $L1-L2$ estimator}
The linear pre-fit can be relatively far from the solution. 
The efficiency of finding a good estimate with the final minimisation 
can be increased by rotating the direction and translating the position around the pre-fitted track. 
The first strategy presented takes the set of linear pre-fits and the set of hits which fall one RMS from the initial track 
to begin the minimisation.
The hit selection used is rather simple, 
so the problem is contaminated by the presence of a large 
number of outliers, i.e. a large number of scattered photons.  
In this particular case, the solution can be found by minimising the $L1-L2$ estimator   
\begin{equation}
\chi^2 = \sum_{i=1}^{n_{hit}} 2 \sqrt{1 + a_i (t^{res}_{i})^2/2} - 2. 
\label{linear}
\end{equation}

\subsubsection{Fit with a $\chi^2$ estimator}
The second strategy presented takes the best solutions coming from the scanning pre-fit procedure and since the hit cleaning 
is particularly efficient, the hit time residuals already follow a Gauss distribution. 
In this particular case, the method of least squares ($\chi^2$)
can be applied to find of the best track parameters
\begin{equation}
\chi^2 = \sum_{i=1}^{n_{hit}} ({t^{res}_{i}})^2.
\label{linear}
\end{equation}
As a result, the strategy gives a set of candidate fits.  
The candidate fits are ordered firstly on the number of hits and secondly on the $\chi^{2}$ normalised to the number of degrees of freedom.

\begin{figure}
\begin{center}
\noindent
\includegraphics [width=0.44\textwidth]{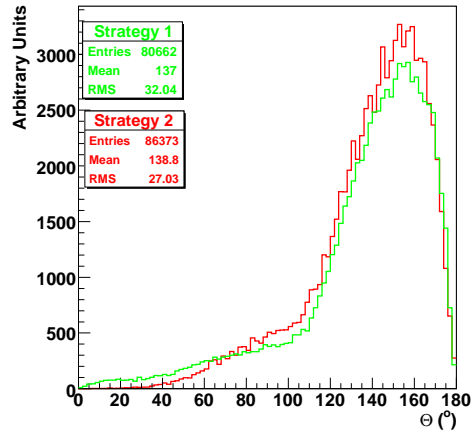}
\end{center}
\caption{\small{Zenith angle distributions given in arbitrary units resulting from the two reconstruction strategies described in the paper.
In green the distribution obtained for the first strategy presented (linear pre-fit and L1-L2 estimator fit) 
and in red the distribution obtained for the second strategy presented (scanning pre-fit and $\chi^2$ fit). 
No analysis cuts are applied.}}\label{zenith}
\end{figure}

\section{Comparison of the two methods}
The two different strategies were applied to the same set of ANTARES 5-line data.
Each event has been carefully calibrated in time, charge and position. 
The first strategy presented has a reconstruction efficiency on triggered events of 85\% while the second one of 91\%. 
The zenith angle distributions resulting from the two reconstruction methods are shown in Fig. \ref{zenith}.
In green the distribution obtained for the first strategy (linear pre-fit and L1-L2 estimator fit) 
and in red the distribution obtained for the second strategy (scanning pre-fit and $\chi^2$ fit). 
No analysis cuts were applied. 
The two different methods presented do not make use of any detailed likelihood description, 
they use approximations with analytical estimators, but these preliminary results obtained are encouraging. 
The two strategies can be improved with the inclusion of a more detailed description of the PDF. 


\bibliography{libros}
\bibliographystyle{unsrt} 

\end{document}